\documentclass[12pt]{article}
\usepackage{amssymb}
\usepackage{euscript}

\usepackage[latin1] {inputenc}
\usepackage[T1]{fontenc}
\usepackage{amsmath}
\usepackage{amsfonts}
\usepackage{amssymb}
\usepackage{latexsym}
\usepackage[english]{babel}
\usepackage[dvips]{graphicx,color}
\usepackage{latexsym}
\usepackage{graphicx}
\usepackage{graphics}
\usepackage{color}
\usepackage{fancyhdr}
\usepackage{fancyhdr}
\usepackage{pstricks}

\newcommand{\be}{\begin{equation}}
\newcommand{\ee}{\end{equation}}
\newcommand{\beq}{\begin{equation}}
\newcommand{\eeq}{\end{equation}}
\newcommand{\bea}{\begin{eqnarray}}
\newcommand{\eea}{\end{eqnarray}}

\newcommand{\ba}{\begin{eqnarray}}
\newcommand{\ea}{\end{eqnarray}}
\def\bc{\begin{center}}
\def\ec{\end{center}}

\begin{document}

\begin{titlepage}

\vspace{2in}

\begin{centering}
{\Large {\bf Area Operators in Holographic Quantum Gravity}}

\vspace{.3in}
Marcelo Botta Cantcheff \footnote{e-mail:
bottac@cern.ch, botta@fisica.unlp.edu.ar}

\vspace{.2 in}

{\it IFLP-CONICET CC 67, 1900,  La Plata, Buenos Aires, Argentina}\\

\vspace{.4in}

\emph{\small{Essay written for the Gravity Research Foundation 2014 Awards for Essays on Gravitation}}

 March 31, 2014

\begin{abstract}
\noindent

We argue that the holographic formula relating entanglement entropy and the area of a minimal surface is the key to \emph{define} the area of surfaces in the (emergent) spacetime from the dual theory on the boundary. So we promote the entropy/area relation to operators to define the \emph{area} observable in a holographic formulation of quantum gravity, then we find a suitable geometric representation for the states, and show that the Ryu-Takayanagi proposal is recovered in the approximation of semiclassical gravity. Finally, we discuss this picture in the example of a AdS-Black hole.

\end{abstract}

\end{centering}

~

\noindent

\end{titlepage}

\vspace{0.7cm}

\textbf{Introduction}

\vspace{0.5cm}

The gauge/gravity correspondence suggests that the spacetime geometry, with certain fixed asymptotic behavior, emerges from the
   degrees of freedom of an ordinary quantum field theory defined on its asymptotic boundary \cite{adscft}. We can view this field theory as the exact (non-perturbative) description of gravity even at quantum level. This is what we refer to as holographic quantum gravity (HQG), although
     we only understand features of this paradigm for semiclassical gravity.  We really do not know how the basic properties of the classical and quantum geometry can be built up from the gauge theory, but the expectation is that quantum entanglement between subsystems of the dual field theory should be a key ingredient in this mechanism \cite{VR, collapse}. One of the most fundamental questions related is \emph{which are} suitable observables of the geometry in a consistent holographic formulation of quantum gravity.


      In this essay, we focus on this problem and argue that the \emph{area} of spatial surfaces in a emergent quantum geometry can be an observable of the theory of gravity through a holographic definition. This is done by promoting to operators the holographic relation entropy/area, very studied in the the context of gauge/gravity correspondence \cite{takaya1,takaya2}. We will show that this can have important consequences to describe the states of HQG in a geometric basis, and also observe that those describing a higher dimensional spacetime are highly entangled, in agreement with the arguments of \cite{VR} and \cite{collapse}.

\vspace{0.7cm}

\textbf{Entanglement entropy and holographic area }

\vspace{0.5cm}

In a local quantum field theory, where the set of degrees of freedom
can be separated in two disjoint subsystems as $\sigma \cup
\bar{\sigma}$, and the Hilbert space as $ {\cal H}_\sigma \otimes
{\cal H}_{\bar{\sigma}} $, the entanglement entropy is defined as
\be\label{def-s} s(\sigma) = - Tr_{\sigma}\, \left(\rho(\sigma) \log
\rho(\sigma) \right) .\ee If the total system is in
      a pure state $|\psi > $,  the reduced density matrix
        $ \rho(\sigma) = Tr_{\bar{\sigma}} |\psi><\psi|$ (normalized to be $ Tr_{\sigma}\,\rho(\sigma) =1$) describes
         the subsystem $\sigma$ in a mixed state. 

In the most familiar example of holography, the AdS/CFT
correspondence, the CFT is
 an ordinary field theory defined on $S^d\times {\mathbb R}$ and its
   degrees of freedom live on the sphere $S^d$. The conjecture is that the states of this theory correspond
to spacetimes with the same asymptotic behavior as global AdS$_{d+2}$ spacetime\footnote{$S^d\times {\mathbb R}$ is the asymptotic boundary of the
global $AdS_{d+2}$ spacetime.}.


   Let $N$ a spacial slice of an asymptotically AdS spacetime $(M, g_{\mu \nu})$ whose induced (Riemannian)
    metric is $h_{ij}$, and $\partial N = S^d$.
   This corresponds to a fixed time in the CFT.

There is a remarkable holographic formula proposed by Ryu and Takayanagi \cite{takaya1, takaya2},
 which has been extensively tested in the last years \cite{review-TR}, that relates the
entanglement entropy (\ref{def-s}) in a strongly coupled quantum field theory and the area of a
special surface in the dual holographic spacetime:
\be\label{TR} s(\sigma) = \frac{1}{4 l^2}\, a_\gamma(\Sigma_{min}) \,\,\,\,\,\,\,\,,\,\,\, \sigma \subset S^d\ee
where
$a_\gamma(\Sigma)$ denotes the area of the $d$-dimensional
hypersurfaces $\Sigma (\subset N)$ homolog to $\sigma$, such that
 $\partial \Sigma = \partial \sigma \equiv \gamma$; and $\Sigma_{min}$ is the one that minimizes this area.
  The constant $l^2\equiv G_N^{(d+2)}$ is a fundamental area of the theory of gravity in $d+2$ dimensions, in the context of AdS/CFT it is related to
  the parameters of the string theory.


This relation should play a essential role in viewing the area as an emergent concept in holographic gravity. So by
 computing an entropy in the field theory, something that depends only on the state, one in principle might obtain
  the area of certain surfaces in the dual spacetime and view this as an \emph{observable} of the emergent geometry.
   Then it would be promoted to operator in some specific formulation of HQG.

\vspace{0.7cm}

Let us notice that the entanglement entropy (\ref{def-s}) can be
expressed as the expectation value of an operator defined such that
\be\label{def-S-operator} \hat{S}(\sigma) = - \log \rho(\sigma)
\,\,,\ee

since \be s(\sigma) = <\psi|\hat{S}(\sigma)|\psi>=  - Tr_{\sigma}\, \left(\rho(\sigma) \hat{S}(\sigma)\right) =
 - Tr_{\sigma}\, \left(\rho(\sigma) \log \rho(\sigma) \right).\ee
So one can name this the (entanglement) \emph{entropy operator} for the reduced system $\sigma$.
Since here we assume that $|\psi\!>$ is the ground state of the whole system, this is what is known in the literature as the modular Hamiltonian \cite{haag}.

\vspace{0.7cm}

\textbf{The area operator and HQG states}

\vspace{0.5cm}

The natural expectation is that in the quantum gravity regime of AdS/CFT,
 the spacetime metric is fluctuating, and thus it is difficult to give a background independent
  definition of the area of surfaces. So then, the right hand
   side of the formula (\ref{TR}) seems to be senseless in a quantum geometry.
One can simply view $\Sigma_{\min}$ in (\ref{TR}) as a particular emergent surface in a holographic
 space whose area is given by the expectation value of the entropy operator, or adopt a less
  conservative point of view, and admit that this operator may be interpreted as holographic dual to
   the area observable in a quantum theory of gravity.

 Thus, if the area is also promoted to an operator of the dual quantum gravity
by means of
\be\label{def-a-operator}
\hat{S}(\sigma) = \frac{1}{4 l^2}\,A_\gamma \, .
\ee
 Let us remark that here $A_\gamma$ stands for an Hermitian operator on the reduced Hilbert space $\cal{ H}_\sigma$,
  which unlike (\ref{TR}), is not referred to \emph{any special surface} in the space.
Let us discuss now the context in which this can make sense.



In the approximation semiclassical of gravity, each state $|\psi> \in {\cal H}$ of the conformal field
 theory on $S^d$ corresponds to a space-like hypersurface $(N, h_{ij})$ of some specific asymptotically
  $AdS_{d+2}$ spacetime, such that $\partial N = S^d$.
So in the quantum gravity regime, each state $|\psi>$ should correspond to a specific \emph{initial} wave
 function of the spatial geometry,
 formally represented as $\psi_g =\psi_g (N, h_{ij}, \phi)$, where the conventional configuration variables
  of quantum gravity are $N, h_{ij}$, and $\phi$ stands for (bulk) matter fields.
For instance, the wave function for the ground state
    can be defined using a path integral approach, by summing over Euclidean compact spacetimes with boundary
     conditions $(N, h_{ij})$ and the same asymptotics as global Euclidean AdS spacetime \cite{HH}. This is the
      Hartle-Hawking construction of the wave function\footnote{This approach was applied to the context of AdS/CFT to describe eternal black holes \cite{eternal}}.

\vspace{0.2cm}

Nevertheless, we will argue here that in HQG, the states can be represented in another suitable basis
 related to the bulk geometry: a set of $d$-dimensional spacelike surfaces $\Sigma \subset N$



\vspace{0.7cm}

By virtue of (\ref{def-S-operator}) and (\ref{def-a-operator}) we have that
the reduced density matrix can be expressed in terms of the area
operator \be\label{rho-area} \rho(\sigma) = C\, e^{- A_\gamma/4 l^2}\,
\, , \ee where $C$ is the normalizing constant.  Remarkably, this is an object entirely referred to the (quantum)
theory of gravity.

Naturally, the factorization $ {\cal H}_\sigma \otimes {\cal H}_{\bar{\sigma}} $ implies that
 the dual $ {\cal H}_{grav}$ splits in some way such that part of the information on the quantum
  state of the spacetime was traced out, and the mixed state (\ref{rho-area}) represents it in a
   \emph{coarse grained} description of quantum gravity. In fact,  (\ref{rho-area})
is what describes the remaining/effective gravitational degrees of freedom.
We will see now that precisely these should be related to a structure of surfaces whose area is quantized.

\vspace{0.3cm}


By virtue of eqs. (\ref{def-S-operator}) and (\ref{def-a-operator}), $A_\gamma $ is
 positive semidefinite, so it can be diagonalized by a orthonormal basis of $\cal{ H}_\sigma$.
  Its eigenvalues (areas by definition) are attributes of certain $d$-dimensional spacelike
   surfaces, and so the spectrum
 shall be \emph{literally} interpreted as the areas of a family
of $d$-dimensional surfaces $\Sigma \subset N$, homolog to $\sigma$ and such
 that $\gamma = \partial \Sigma$.
Thus for a given spatial topology $N$, the area operator can take a definite value for states
 associated to such surfaces, say
\be
A_\gamma \, |\Sigma_i, \dots> =\, a (\Sigma_i) \,\,|\Sigma_i, \dots> \, .
\ee
So $A_\gamma$ has a spectral decomposition such that
  $A_\gamma = \sum_{i}\, a (\Sigma_i) |(\Sigma ,\dots)_i\rangle\langle(\Sigma , \dots)_i| $
   where  $|(\Sigma , \dots )_i\rangle$ are orthonormal states in $\cal{ H}_\sigma$, and $\dots$
    denote the set of quantum numbers that characterize completely each surface $i$ in order to
     have a complete nondegenerate spectrum of states; for instance, since $\sigma$ is homolog to
      $ \Sigma_i$, one can specify each of these surfaces
       by giving an embedding field ${\bf z}_i\,:\,\sigma \mapsto  \Sigma_i \left(\subset N\right)$,
        allowing the pullback of extra geometrical structures; in particular, it would provide a
         $d \times d$ Riemannian metric on each $\Sigma_i$.
   The index $i$ labels a basis of  eigenstates of the area operator, and it might be continuum
   or discrete depending on the structure of the Hilbert space of the dual field theory
    (eq. (\ref{def-S-operator})). Below we discuss more this point in the example of an AdS black hole.

Near the boundary $\sigma$, the manifold
 $N$ can always be described as a foliation in a one-parameter family of
surfaces homolog to $\sigma$; then in this sense, if the quantum spectrum
of areas is discrete, one can view it as \emph{approximating} certain
foliation of $N$.

Then, if $\{(\Sigma_i , \dots)\}$ form a complete basis for the reduced Hilbert space, the HQG general wave functions shall be functionals of these surfaces $\psi= \psi(\Sigma_i , \dots)$ and the
 mean value of the area operator expresses as,
\be
<\psi | A_\gamma |\psi> = \sum_i \,a (\Sigma_i) \, |\psi(\Sigma_i , \dots)|^2 \, ,
\ee
for an arbitrary state $|\psi\!\!> \in {\cal H}_\sigma$. This suggests that co-dimension $2$ surfaces can constitute configuration variables more convenient for holographic gravity.

By virtue of the formula (\ref{rho-area}), one can represent the (coarse-grained) \emph{state of the spacetime}, as a probability distribution
 on geometries labeled by surfaces $\Sigma_i \subset N \;$:
 \be
\rho(\sigma)= \sum_i \, e^{- a(\Sigma_i)/4 l^2}\, |(\Sigma_i ,\dots)\rangle\langle(\Sigma_i ,\dots)|\, . \ee
A \emph{purification} of this is a state $| \psi_\gamma\rangle \in {\cal H}_\sigma \otimes {\cal H}_{\bar{\sigma}}$ such that
 \be
\rho(\sigma)= Tr_{\bar{\sigma}} \left( | \psi_\gamma\rangle \langle \psi_\gamma| \right)
 \ee
Since a similar analysis should be possible for the complement  $\bar{\sigma}$,
we have that
\be
\rho(\bar{\sigma})= \sum_i \, e^{- a(\bar{\Sigma}_i)/4 l^2}\, |(\bar{\Sigma}_i ,\dots)\rangle\langle(\bar{\Sigma}_i ,\dots)| .\ee
 In particular if we choose $\sigma=\bar{\sigma}$, the respective spectra of surfaces shall be coincident and the pure state
 writes
  in geometrical variables as:
 \be\label{state-surfaces}
| \psi_\gamma\rangle= \sum_i \, e^{- a(\Sigma_i)/8 l^2}\, |(\Sigma_i ,\dots)\rangle \otimes |(\bar{\Sigma}_i ,\dots)\rangle \, = \sum_i \, e^{- a(\bar{\Sigma}_i)/8 l^2}\, |(\Sigma_i ,\dots)\rangle \otimes |(\bar{\Sigma}_i ,\dots)\rangle \,\, \in {\cal H}_{grav}
 \ee
 is the quantum state of the spacetime written in geometric variables\footnote{This expression resembles the general proposal done on Ref. \cite{collapse}}.

 This state can literally be interpreted as a \emph{quantum superposition} of (geometric) states
  characterized by surfaces of area $a$, or simply as a quantum superposition of surfaces in $N$.

\newpage

\textbf{The minimal surface proposal}

\vspace{0.5cm}

Let us now show that our quantization hypothesis (\ref{def-a-operator}) is consistent with the Riu-Takayanagi prescription (\ref{TR}). In fact
 the minimal surface proposal can be derived only by assuming that the entropy/area can be promoted to operators
  according to the rule (\ref{def-a-operator}).

In fact, by taking the expectation value of (\ref{def-a-operator}), and  using (\ref{def-S-operator}) and (\ref{def-a-operator}) we obtain
\be
 \langle \,\hat{S}(\sigma) \,\rangle=  \frac{1}{4}\, \left \langle\, A_\gamma \,\right \rangle = \frac{1}{4l^2}\,Tr_\sigma\, \rho(\sigma) \,A_\gamma = \frac{C}{4l^2}\,Tr_\sigma\, e^{- \,A_\gamma/4 l^2}\,\,\,A_\gamma
\ee
where $C^{-1}\equiv Tr_\sigma\, e^{- \,A_\gamma/4 l^2}$.
If we express it in a basis such that $A_\gamma$ is diagonal, say $\{|\Sigma_i \dots> \,/\, \partial \Sigma_i =\gamma \}$, results
\be
s(\sigma) = \langle \hat{S}(\sigma) \rangle= \frac{C}{4 l^2 }\,\sum_{i} \,  e^{- a(\Sigma_i)/4 l^2}\,\,a(\Sigma_i)
\ee
From this expression it can be seen that the dominant term in this sum is the one with the minimum value of the area.
So in a standard saddle point approximation for $l^2 \ll a (\Sigma_i) \,\, (\forall i$) we recover
\be
s(\sigma) \approx \frac{C}{4l^2 }\,  e^{- a(\Sigma_{min})/4 l^2}\,\,a(\Sigma_{min}) \,\approx \frac{1}{4l^2 }\,  \,a(\Sigma_{min})\,\,\,,
\ee
where we have used that $C^{-1} \approx e^{- a(\Sigma_{min})/4 l^2}\,$.

\vspace{0.5cm}

\textbf{Entaglement, mixed states, and the emergent space}

\vspace{0.5cm}

\begin{figure}
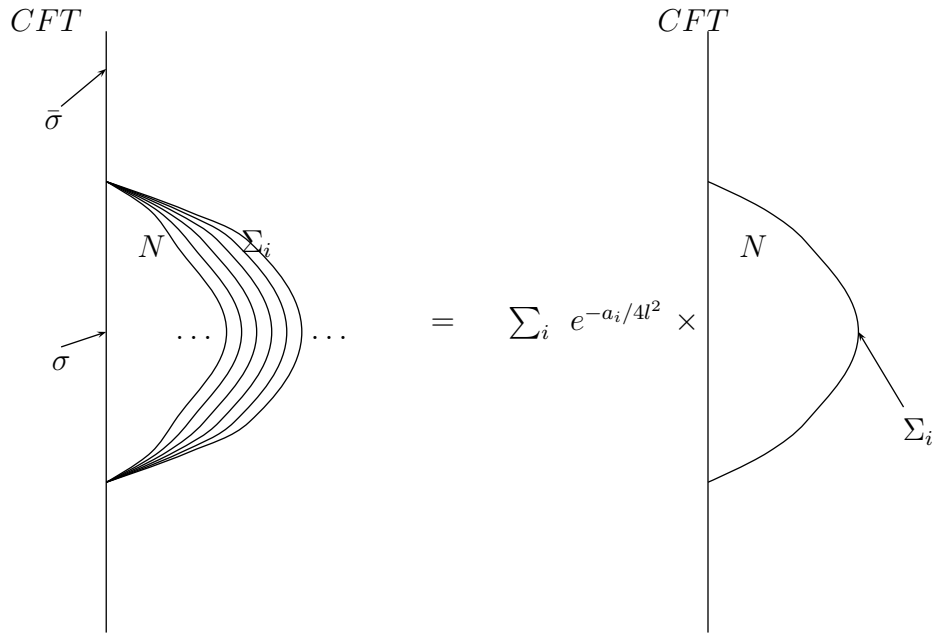

\hspace{3cm}
\pscurve[linewidth=0.5pt](1,-6)(2.1,-6.4)(2.9,-6.8)(3.6,-8)(2.9,-9.2)(2.1,-9.6)(1,-10)
\pscurve[linewidth=0.5pt](1,-6)(2,-6.4)(2.7,-6.8)(3.4,-8)(2.7,-9.2)(2,-9.6)(1,-10)
\pscurve[linewidth=0.5pt](1,-6)(1.9,-6.4)(2.5,-6.8)(3.2,-8)(2.5,-9.2)(1.9,-9.6)(1,-10)
\pscurve[linewidth=0.5pt](1,-6)(1.8,-6.4)(2.3,-6.8)(3,-8)(2.3,-9.2)(1.8,-9.6)(1,-10)
\pscurve[linewidth=0.5pt](1,-6)(1.7,-6.4)(2.1,-6.8)(2.8,-8)(2.1,-9.2)(1.7,-9.6)(1,-10)
\pscurve[linewidth=0.5pt](1,-6)(1.6,-6.4)(1.9,-6.8)(2.6,-8)(1.9,-9.2)(1.6,-9.6)(1,-10)
\pscurve[linewidth=0.5pt](9,-6)(9.8,-6.4)(10.3,-6.8)(11,-8)(10.3,-9.2)(9.8,-9.6)(9,-10)
\psline[linewidth=.5pt]{<-}(1,-8)(0.4,-8.2)
\rput[bI](0.4,-8.5){$\sigma$}
\psline[linewidth=.5pt]{<-}(1,-4.5)(0.4,-5)
\rput[bI](0.3,-5.3){$\bar{\sigma}$}
\rput[bI](7.1,-8.1){$=\;\;\;\;\;\sum_i \,\; e^{- a_i/4 l^2}\,\, \times$}
\rput[bI](4,-8.1){$\dots$}
\rput[bI](2.2,-8.1){$\dots$}
\psline[linewidth=0.5pt](1,-4)(1,-12)
\psline[linewidth=0.5pt](9,-4)(9,-12)
\rput[bI](0.2,-4){$CFT$}
\rput[bI](8.8,-4){$CFT$}
\rput[bI](3,-7){$\Sigma_i$}
\rput[bI](9.6,-7){$N$}
\rput[bI](1.6,-7){$N$}
\rput[bI](11.8,-9.5){$\Sigma_i$}
\psline[linewidth=.5pt]{<-}(11,-8)(11.6,-9)
\vspace{12cm}
\caption{\small{This shows schematically the state (\ref{state-surfaces}), that represents the space $N$ as quantum superposition of surface states.}}
\label{figure}
\end{figure}

Let us notice the similarity of (\ref{state-surfaces}) with the
thermal density matrix in a canonical ensemble. So the simplest
interpretation of $\rho(\sigma)$ is viewing this as a mixed state
with a distribution of probability
 given in terms of the observable $A_\gamma$ in HQG. In this way, the $d+1$-dimensional 'space'
$N$ emerges as a distribution of surfaces, in a similar way to a
distribution of particles among the states of gas. Thus $e^{-a/4
l^2}$ is the Boltzman-like probability, or statistic frequency,
 of having a surface of area $a$ in the space $N$.
  In fact $N$ (foliated in $\Sigma$`s) should be thought as an ordinary
 macroscopic system in a mixed state,
  which is probed  through
 a large number of independent observations (\emph{ensemble}) in the
 space-time (see Fig. \ref{figure}).



Moreover, this result emphasizes that entanglement is fundamentally related to the emergence of the space(time); in fact, let us notice that if the entanglement entropy vanishes ($s(\sigma)=0$), then the entropy/area operator is totally degenerate (the eigenvalues are $0$ or $1$) and so there is \emph{only one} emergent surface state $\Sigma \subset N$ \footnote{Or many with coincident areas.} rather than a family or foliation $\{\Sigma\}_i \approx N$ filling the space. Thus, in the sense discussed here, one would not have a holographic $d+1$-dimensional emergent space.

 We will see below that the cited connection with a canonical ensemble is clear in the context of black holes, where the energy spectrum of CFT can
  be explicitly related to the spectrum of surfaces in the gravity side.

\newpage

\textbf{Thermal states and black holes}

\vspace{0.5cm}

Thermofield dynamics (TFD) \cite{tu, ume} seems to be the most appropriate framework to describe the dual holographic
 of eternal (AdS) black holes \cite{eternal}, and any spacetime with two asymptotic AdS regions causally disconnected \cite{collapse},
  which has thermodynamic properties. This is a well understood context where entanglement play a central role and
   we have a clear interpretation of what the dual spacetime is \cite{eternal}, so most of the ideas and results discussed previously
  can be tested and interpreted more clearly.


In the TFD formalism the statistical properties of a system are described by pure states in
${\cal H}\otimes\widetilde{{\cal H}}$, where ${\cal H}$ is the Hilbert space of the accessible subsystem and $\widetilde{{\cal H}}$
 is a (decoupled) copy referred to as the \emph{thermofield double}. So the thermal equilibrium corresponds to the entangled state
 \be\label{state-tfd}
\left.\left|\psi(\beta)\right\rangle \! \right\rangle = Z^{-1} \,\sum_{n}  e^{- \beta E_n/2}\,\left|n\right\rangle \, \otimes\,\left| \tilde{n} \right\rangle
\ee
where $n$
represents the \emph{n}th energy eigenstate of the Hamiltonian $H$, and its copy $\tilde{n}$ spans $\widetilde{{\cal H}}$.
By tracing $ \left|\psi(\beta)\right\rangle \left\langle \psi(\beta) \right|$ over $\widetilde{{\cal H}}$ one indeed recovers the conventional
thermal density operator $\rho(\beta) = Z^{-1} \, e^{- \beta \hat{H}/2}$.

 In TFD, the thermodynamic entropy is regarded as \emph{entanglement entropy} between both sectors,
  and it is defined as \emph{operator}.
   For instance, in a free quantum field theory this operator can be
expressed \emph{a priori} in terms of the canonical occupation number operator $\hat{N}_{n}$ of the mode $n$
as \cite{tu}:
\begin{equation}
\hat{S} = \,-\, \sum_{n}\left[\hat{N}_{n}\log \left( \hat{N}_{n}\right) -
\left( 1+\hat{N}_{n}\right) \log \left( 1+\hat{N}_{n}\right)  \right] \,.
\label{entrboson}
\end{equation}
For a thermal distribution $\hat{N}_{n}$ is a function of $\beta$ (see refs \cite{tu, ume} for details); and in this case, it is not difficult to show that the state (\ref{state-tfd}) can be written in terms of this operator as \footnote{A detailed demonstration of this expression may be found in Refs. \cite{vitiello}.}
\be\label{state-tfd-ent}
\left.\left| \psi (\beta)\right\rangle \! \right\rangle =  e^{- \hat{S}(\beta)/2} \sum_{n } \left|n\right\rangle \otimes \left| \tilde{n} \right\rangle\,.
\ee
Then, the reduced density matrix results
\be\label{rho-beta}
\rho(\beta) \equiv Tr_{\widetilde{{\cal H}}} \left|\psi(\beta)\right\rangle \left\langle \psi(\beta) \right|= e^{- \hat{S}(\beta)} \,\,,
\ee
 which indeed is consistent with the expression (\ref{def-S-operator}).


On the other hand, if the system is CFT and $\sigma \equiv S^d$, the complement/TFD-double $\bar{\sigma}$ is another
 \emph{disconnected} copy of $S^d$. In this case we know that the state (\ref{state-tfd}) is \emph{literally dual} to the maximally extended
 AdS-Schwarschild black hole \cite{eternal}.



Thus, according to the holographic proposal (\ref{def-a-operator}) we may write
\be\label{rho-beta-area}
\rho(\beta) =  C \,e^{- A/4 l^2}
\ee
where here $A$ has a spectrum of surfaces $\Sigma_i$ homolog to $S^d$, and so by the same arguments followed in the
 previa sections we obtain that the state can also be expressed as (\ref{state-surfaces}).
Moreover, if we compute as before the mean value of $A$ for this distribution in the semiclassical gravity approximation, results that
 the observed entropy of the black hole is
\be
s(\beta) \approx  a(\Sigma_{min})/4 l^2 .
\ee
This is consistent with the Bekenstein-Hawking law remembering that, in a static black hole, the surface
  $\Sigma_{min}$ ($ \partial\Sigma_{min} = \partial\sigma$) wraps around the horizon as $\sigma \to S^d$, and so
   the regularized area coincides with that of the event horizon \cite{Azeyanagi}. In this sense one might observe that
    there is no surfaces $\Sigma_i \,\,( \sim S^d)$ probing the space behind the horizon because $a(\Sigma_i) \geq a_{min}\,\,\, $  \cite{mathur}

   Furthermore, if the CFT Hamiltonian commutes with $\hat{S}$ and the proposal (\ref{def-a-operator}) is correct, the expression (\ref{rho-beta-area}) (or (\ref{state-surfaces})) coincides exactly with $\rho(\beta) = Z^{-1} \, e^{- \beta \hat{H}/2}$ (or (\ref{state-tfd})), and thus
     we have a remarkable interpretation for each CFT energy eigenstate  $|n\!>$ : as \emph{dual to a surface state $(\Sigma, \dots)_{n} $ in the gravity side}.

In the weak coupling limit of the dual field theory, the state (\ref{state-tfd}) supposedly describes a dual spacetime in the quantum gravity regime,
  that one might interpret as a quantum black hole. In this case, because of the compactness of $S^d$, the energy
   modes $n$ are discrete and by virtue of the formula (\ref{entrboson}) (and (\ref{def-a-operator})), the entropy/area of the $\Sigma_n$'s $( \sim S^d)$ are \emph{quantized}.



\vspace{0.7cm}

\textbf{Conclusions}

\vspace{0.5cm}

In this essay, we have defined holographically an area operator involving the entanglement entropy between subsystems of the dual theory in the boundary, and it was done consistently with the holographic recipe \cite{takaya1} in the classical limit.

We have argued that this simple assumption leads to a specific form of the (mixed) state of the spacetime in terms of a geometrical representation related to co-dimension two surfaces. This might be a suitable configuration basis for the Hilbert space of HQG that shall be more investigated.

It is fascinating that this pictures manifestly captures the significance of the quantum entanglement in the holographic emergence of the spacetime, in agreement with arguments previously presented \cite{VR} \cite{collapse}.

Let us mention finally that the idea of quantizing the area has also been developed in the context of \emph{loop quantum gravity}, but the spirit and construction presented here are completely different.\\

\vspace{0.7cm}

{\small \textbf{Acknowledgements}
 The author is grateful to Raul Arias, A. Gadelha, N. Grandi and G. Silva for useful discussions. This work was partially supported by: CONICET PIP 2010-0396.}

\end{document}